\documentclass[lettersize,journal]{IEEEtran}
\usepackage{cite}

\usepackage[english]{babel}

\usepackage{amsmath,amssymb,amsfonts}
\usepackage{algorithmic}
\usepackage{graphicx}
\usepackage{textcomp}
\usepackage{bm}
\usepackage{tikz}
\usepackage{multirow}
\usepackage{balance}

\DeclareMathAlphabet{\mathscrbf}{OMS}{mdugm}{b}{n}

\usepackage[binary-units]{siunitx}

\DeclareSIUnit{\mT}{\milli\tesla}
\DeclareSIUnit{\kHz}{\kilo\hertz}
\DeclareSIUnit{\MHz}{\mega\hertz}
\DeclareSIUnit{\Tpm}{\tesla\per\metre}
\DeclareSIUnit{\mul}{\micro\litre}
\DeclareSIUnit{\mgml}{\milli\gram/\milli\litre}
\DeclareSIUnit{\cm}{\centi\metre}
\DeclareSIUnit{\mm}{\milli\metre}
\DeclareSIUnit{\mum}{\micro\metre}
\DeclareSIUnit{\mumol}{\micro\mol}
\DeclareSIUnit{\degC}{\celsius}

\newcommand{\N}{\mathbb{N}}

\newcommand{\R}{\mathbb{R}}
\newcommand{\C}{\mathbb{C}}

\newcommand{\eUnit}{\mathrm{e}}
\newcommand{\iUnit}{\mathrm{i}}
\newcommand{\mo}{m_0}

\newcommand{\kB}{k_\mathrm{B}}
\newcommand{\TP}{T_\mathrm{P}}
\newcommand{\Vc}{V_\mathrm{c}}

\newcommand{\Kanis}{K^\text{anis}}
\newcommand{\Kanismax}{K^\text{anis}_\text{max}}

\newcommand{\dt}[1][t]{\,\mathrm{d}#1}

\newcommand{\mbar}[1]{\bar{\bm{m}}\!\left(#1\right)}
\newcommand{\tran}{\intercal}
\newcommand{\ak}{\alpha_\mathrm{K}}
\newcommand{\HD}{\bm{H}^\mathrm{D}}
\newcommand{\HS}{\bm{H}^\mathrm{S}}
\newcommand{\TD}{T_\mathrm{D}}
\newcommand{\NB}{N_\mathrm{B}}
\newcommand{\fB}{f_\mathrm{B}}
\newcommand{\rect}{\operatorname{rect}}
\newcommand{\sgn}{\operatorname{sgn}}

\definecolor{ibilight}{RGB}{193,216,237}
\definecolor{ibidark}{RGB}{0,73,146}	
\definecolor{uke2}{RGB}{170,156,143} 	
\definecolor{uke3}{RGB}{87,87,86}		
\definecolor{ukesec1}{RGB}{255,223,0}	
\definecolor{ukesec2}{RGB}{239,123,5}	
\definecolor{ukesec3}{RGB}{104,195,205}	
\definecolor{ukesec4}{RGB}{138,189,36}	
\definecolor{ukesec5}{RGB}{178,34,41}	
\definecolor{tuhh}{RGB}{45,198,214}     
\definecolor{ibidarkBG}{RGB}{227,229,242}   
\definecolor{uke2BG}{RGB}{233,228,225} 	    
\definecolor{uke3BG}{RGB}{230,231,232}	    
\definecolor{ukesec1BG}{RGB}{255,243,190}   
\definecolor{ukesec2BG}{RGB}{254,232,212}   
\definecolor{ukesec3BG}{RGB}{222,241,241}   
\definecolor{ukesec4BG}{RGB}{233,243,222}   
\definecolor{ukesec5BG}{RGB}{244,230,225}   

\def\BibTeX{{\rm B\kern-.05em{\sc i\kern-.025em b}\kern-.08em
    T\kern-.1667em\lower.7ex\hbox{E}\kern-.125emX}}
\begin{document}
\title{Efficient Chebyshev Reconstruction for \\ the Anisotropic Equilibrium Model \\ in Magnetic Particle Imaging}
\author{ Christine~Droigk, Daniel~H.~Durán, Marco~Maass,~\IEEEmembership{Member,~IEEE}, Tobias~Knopp, and Konrad~Scheffler
\thanks{
Submitted on Month Day, 2025. D. H. Durán acknowledges DASHH, Data Science in Hamburg - Helmholtz Graduate School for the Structure of Matter, for financial support.
}
\thanks{
Christine Droigk is with the Institute for Signal Processing, University of Luebeck, 23562 Luebeck, Germany (e-mail: c.droigk@uni-luebeck.de).
}
\thanks{
Daniel H. Durán is with the Institute for Biomedical Imaging, Hamburg University of Technology, 21073, Hamburg, Germany and the Centre for X-ray and Nano Science CXNS, Deutsches Elektronen-Synchrotron DESY, 22607
Hamburg, Germany 
(e-mail: daniel.hernandez.duran@tuhh.de).
}
\thanks{
Marco Maass is with the German Research Center for Artificial Intelligence (DFKI), 23562 Luebeck, Germany, and also with the Institute of Medical Informatics, University of Luebeck, 23562 Luebeck, Germany (e-mail: marco.maass@dfki.de).
}
\thanks{
Tobias Knopp is with the Section for Biomedical Imaging, University Medical Center Hamburg-Eppendorf, 20246 Hamburg, Germany, also with the Institute for Biomedical Imaging, Hamburg University of Technology, 21073 Hamburg, Germany, and also with the Fraunhofer Research Institution for Individualized and Cell-Based Medical Engineering IMTE, 23562 Luebeck, Germany (e-mail:
t.knopp@uke.de).
}
\thanks{
Konrad Scheffler is with the Section for Biomedical Imaging, University
Medical Center Hamburg-Eppendorf, 20246 Hamburg, Germany, and also with
the Institute for Biomedical Imaging, Hamburg University of Technology, 21073 Hamburg, Germany (email: konrad.scheffler@tuhh.de).
}
}

\maketitle

\begin{abstract}
Magnetic Particle Imaging (MPI) is a tomographic imaging modality capable of real-time, high-sensitivity mapping of superparamagnetic iron oxide nanoparticles. Model-based image reconstruction provides an alternative to conventional methods that rely on a measured system matrix, eliminating the need for laborious calibration measurements. Nevertheless, model-based approaches must account for the complexities of the imaging chain to maintain high image quality. A recently proposed direct reconstruction method leverages weighted Chebyshev polynomials in the frequency domain, removing the need for a simulated system matrix. However, the underlying model neglects key physical effects, such as nanoparticle anisotropy, leading to distortions in reconstructed images. To mitigate these artifacts, an adapted direct Chebyshev reconstruction (DCR) method incorporates a spatially variant deconvolution step, significantly improving reconstruction accuracy at the cost of increased computational demands. In this work, we evaluate the adapted DCR on six experimental phantoms, demonstrating enhanced reconstruction quality in real measurements and achieving image fidelity comparable to or exceeding that of simulated system matrix reconstruction. Furthermore, we introduce an efficient approximation for the spatially variable deconvolution, reducing both runtime and memory consumption while maintaining accuracy. This method achieves computational complexity of $\mathcal{O}(N \log N)$, making it particularly beneficial for high-resolution and three-dimensional imaging. Our results highlight the potential of the adapted DCR approach for improving model-based MPI reconstruction in practical applications.
\end{abstract}

\begin{IEEEkeywords}
Anisotropic Equilibrium Model, Efficient Spatially Variant Deconvolution, Chebyshev Reconstruction, Magnetic Particle Imaging, Model-Based Image Reconstruction
\end{IEEEkeywords}

\section{Introduction}
\IEEEPARstart{M}{agnetic} Particle Imaging (MPI) is a tomographic medical imaging modality capable of determining the spatiotemporal concentration distribution of superparamagnetic iron oxide nanoparticles (SPIOs) \cite{Gleich2005}. While still in the preclinical stage, MPI has shown promise for various clinical applications, including vascular imaging \cite{herz2017magnetic,kaul2018magnetic,vogel2020superspeed, tong2023atherosclerosis}, perfusion imaging \cite{ludewig2017magnetic,molwitz2019first,szwargulski2020monitoring}, cancer imaging \cite{zhu2019quantitative,huang2023glioblastoma}, and interventional imaging \cite{salamon2016magnetic,rahmer2017interactive,herz2019magnetic}.
MPI operates by utilizing spatially and dynamically varying magnetic fields to manipulate the magnetic moments of SPIOs, thereby encoding spatial and temporal information into a measurable voltage signal. Reconstructing the SPIO-distribution from this signal poses an ill-posed inverse problem \cite{maerz2015modelbased,erb2018mathematical,Kluth2018a}, for which different approaches have been developed, depending on the excitation sequence used.
Reconstruction methods can broadly be classified into system-matrix-based and direct reconstruction approaches. System-matrix-based methods rely on either a measured or model-based system response of an MPI scanner. The reconstruction problem is typically formulated as a system of linear equations and solved using regularized least squares techniques\cite{Knopp2010PhysMedBio,Storath2017,Bathke2017}. Direct reconstruction approaches, on the other hand, can be categorized into time-domain methods, commonly known as $x$-space MPI \cite{Goodwill2010}, and frequency-domain methods, such as Chebyshev reconstruction \cite{rahmer2009signal,droigk2022direct}. For one-dimensional MPI excitation, the mathematical equivalence between $x$-space and Chebyshev reconstruction has been established \cite{gruettner2013ontheformulation}. While $x$-space MPI enables reasonable image reconstruction for one-dimensional excitation sequences \cite{Goodwill2011,croft2012relaxation}, no real-world reconstructions have been demonstrated for more complex excitation patterns, such as Lissajous field-free point (FFP) trajectories. This limitation may stem from the simplifying assumptions of $x$-space MPI, which become insufficient for higher-dimensional excitations. However, simulation studies suggest that $x$-space reconstruction may be applicable to more complex trajectories \cite{Ozaslan2019}. 
During the preparation of this article, Sanders et al. published a decomposition of the temporal forward MPI system equation into several linear operators that can be calibrated with minimal effort for an MPI scanner by measuring a system delay parameter\cite{Sanders2025}. For this purpose, the system equation has been decomposed into different scanner and particle-dependent linear operators so that it can be efficiently computed numerically. Since the proposed model is physically more complete than the classical $x$-space model, the authors were able to show a superior image reconstruction for one-dimensional excitation in real-world experiments. Although the article shows results for multidimensional excitations, these particular results come only from a simulation study, so the question remains whether the adapted model of Sanders et al. can also be successfully applied to real-world measurements with multidimensional excitations.
In contrast, Chebyshev reconstruction has been successfully applied to real-world measurements with multidimensional excitations \cite{droigk2022direct}.

Recently, a more advanced model-based approach incorporating magnetic anisotropy in equilibrium has been proposed~\cite{maass2023system,maass2024equilibrium}. This extended model has also been applied to the direct Chebyshev reconstruction (DCR), leading to the development of an adapted version (DCR-EQANIS) \cite{droigk2023adaption}. Compared to the conventional DCR (DCR-EQ) that relies on the equilibrium model without anisotropies, DCR-EQANIS demonstrated superior reconstruction accuracy on simulated data in \cite{droigk2022direct} by accounting for magnetic anisotropies. However, this improvement comes at the cost of increased computational complexity due to the need for a spatially varying deconvolution step.

In this work, we provide a comprehensive derivation and detailed exposition of the DCR-EQANIS method, which was previously only presented in an abbreviated form as part of an in-proceedings publication~\cite{droigk2023adaption}. Beyond evaluating its performance on simulated data, we present for the first time an assessment of DCR-EQANIS using experimental measurement data. Specifically, we compare the reconstruction quality of the classical DCR-EQ and the adapted DCR-EQANIS across six different phantoms. Furthermore, we introduce an efficient approximation method for the spatially varying convolution, allowing for a tunable trade-off between accuracy and computational efficiency. This enables the extended convolution operation and its adjoint operator to be computed in ${\mathcal O}(N \log N)$. Furthermore, the convolution kernels can be stored compactly, and unlike in Droigk et al.~\cite{droigk2023adaption}, no explicit convolution matrix needs to be constructed for deconvolution, significantly reducing memory requirements. 
To efficiently apply these computational improvements, we employ an iterative optimization algorithm that enables fast operator evaluations during deconvolution. This approach also allows for the integration of alternative regularization techniques beyond standard Tikhonov regularization, providing greater flexibility in reconstruction.
While this method was previously introduced only in an abbreviated form as part of an in-proceedings publication~\cite{Duran2025}, this work provides a more detailed description of the method, including an in-depth analysis of error behavior, a rigorous evaluation of the reconstruction quality, and a detailed assessment of both computational and memory efficiency.

\section{Methods}
In the following, we introduce the models for the mean magnetic moment that form the foundation of the model-based reconstruction. We then outline the necessary adaptations to integrate this model into the DCR. Finally, we present an approximation method designed to accelerate the deconvolution step in the adapted DCR.

\subsection{Equilibrium Model with Anisotropy}
In \cite{maass2022analytical}, a series expansion for an equilibrium model was presented, enabling the efficient simulation of the mean magnetic moments of SPIOs with  uniaxial anisotropy. This model was investigated both numerically and experimentally in the recently published article \cite{maass2024equilibrium}. 
The equilibrium model with anisotropy (EQANIS) is based on the Stoner-Wohlfarth model, which represents uniaxial SPIO anisotropy using an easy axis and an anisotropy constant. 
In this work, we employ a modified and approximated SPIO model for fluid SPIO tracers, as described in \cite{Kluth2019}. The SPIO parameters are expressed as parametric variants dependent on the spatial variable $\bm x\in\R^3$, namely the easy axis $\bm{n}:\R^3\to\mathbb S^2$ and the anisotropy constant $\Kanis:\R^3\to\R$, both of which vary with the location $\bm x$. Here, $\mathbb S^2\subset \R^3$ denotes the surface of the unit sphere. 
The anisotropy strength is defined as $\ak(\bm x) = \tfrac{\Vc \Kanis(\bm x) }{\kB \TP}$, where $\Vc$ represents the SPIO core volume, $\kB$ is the Boltzmann constant, and $\TP$ is the SPIO temperature.
To simplify the following equations, we consolidate the free parameters into the set of observed parameters $\mathbb O(\bm x) = \{\ak(\bm x),\bm{n}(\bm x)\}$. The model used follows the formulation presented in \cite[Eq. (27)]{maass2024equilibrium}. 
The mean magnetic moment of the SPIOs in the EQANIS model, in the presence of an applied magnetic field $\bm H:\R^3\times \R\to\R^3$, is given by 
\begin{equation}
	\mbar{ \bm{H}(\bm x,t);\mathbb O(\bm x) } 	= \mo \mathscrbf{E}(\beta  \bm{H}(\bm x,t); \mathbb O(\bm x) )
	\end{equation}
	with \begin{equation}
            \mathscrbf{E}(\bm \xi; \mathbb O(\bm x) ) = \nabla_{\bm \xi}\ln\left(\mathcal Z(\bm \xi;\mathbb O(\bm x))\right)
           \end{equation}
            and
            \begin{equation}
	\mathcal Z(\bm \xi;\mathbb O(\bm x))  = \int_{\mathbb{S}^2} \eUnit^{ {\bm \xi}^\tran \bm{m} +\ak(\bm x) \bigl((\bm{n}(\bm x))^\tran \bm{m}\bigr)^2} \dt[\bm{m}],
\end{equation}
where $t\in\R$ denotes the temporal variable, and $m_0$ is the magnitude of the magnetic moment of a single nanoparticle.
The applied magnetic field is modeled as a superposition ${\bm H(\bm x,t) = \HS(\bm x)+\HD(t)}$, where $\HS:\R^3\to\R^3$ represents the static selection field, and $\HD:\R\to\R^3$ denotes the time-dependent drive field.  A common assumption is that the selection field is a linear gradient field, given by ${\HS(\bm x) = \bm G \bm x}$, where ${\bm G=\operatorname{diag}(G_x,G_y,G_z)}$ with ${G_x,G_y,G_z\in\R\backslash\{0\}}$. This formulation corresponds to an FFP MPI scanner. 

The vector-valued MPI system equation is given by 
\begin{equation}
    \bm u(t) = \int_{\mathbb{R}^3} \bm s(\bm x, t) c(\bm x) \dt[\bm x],
		\label{eq:timeDomainSysEq}
\end{equation}
where $\bm u: \R \rightarrow \R^3$ represents the measured voltage signals, ${c:\mathbb{R}^3\rightarrow\mathbb{R}}$ describes the magnetic particle distribution, and ${\bm s:\mathbb{R}^3\times\mathbb{R}\rightarrow\mathbb{R}^3}$ denotes the system function, which can be explicitly determined using the magnetization model.

Accordingly, the $k$-th Fourier series component of the system equation in \eqref{eq:timeDomainSysEq} can be formulated as 
\begin{equation}
    \bm u_k = \int_{\mathbb{R}^3} \bm s_k(\bm x) c(\bm x) \dt[\bm x],
		\label{eq:FreqDomainSysEq}
\end{equation}
where $\bm u_k\in\mathbb{C}^3$ denotes the $k$-th frequency component of the voltage signal vector $\bm u(t)$, and $\bm s_k:\mathbb{R}^3\to\mathbb{C}^3$ represents the $k$-th frequency component of the system function $\bm s(\bm x, t)$.

Given a two-dimensional Lissajous FFP trajectory with period length $\TD>0$, the component-wise Fourier series components $\bm s_k$ for $k\in\mathbb{Z}$ can be given as~\cite{maass2023system, maass2024equilibrium}
\begin{equation}
	\begin{aligned}
		\bm s_{k}(\bm{x}) &=   \bm{M}_{ k} \!\!\int_{\mathbb R^3} \left[\frac{\partial^2}{\partial z_1 \partial z_2}\mathscrbf{E}(\beta \bm{G} \bm z; \mathbb O(\bm x))\right]_{\scriptscriptstyle \bm z = \bm x- \bm y}\\ 
        &\hspace{4.5cm}  \times P^{(2)}_k(\bm{y}) \dt[\bm{y}].
	\end{aligned}
	\label{eq:sysFunc2DCheb}
\end{equation}
Here, the function $P^{(2)}_k: \R^3 \rightarrow \R$ consists of a series of tensor products of weighted Chebyshev polynomials of the second kind. Furthermore, $\omega_k$ is defined as ${\omega_k = \frac{2\pi k}{\TD}}$ and the coefficient matrix ${\bm{M}_k\in\C^{3\times 3}}$ is given by ${\bm{M}_{ k} =-\iUnit \omega_k \mu_0 \mo \operatorname{diag}\bigl(\bm{\hat{a}}(\omega_k)\bigr)\bm{P} }$, where $\mu_0$ is the vacuum permeability,  ${\bm P\in\mathbb{R}^{3\times3}}$ denotes the homogeneous receive coil sensitivities and ${\bm{\hat{a}}:\R^3\to\C^3}$ represents the Fourier transform of the impulse response generated by each receive chain, commonly referred to as the transfer function. 

\subsection{The Reduced EQANIS Model}
The reduced equilibrium model without anisotropy serves as the foundation for the reconstruction method presented in~\cite{droigk2022direct}. However, it can also be formulated for the EQANIS model~\cite{droigk2023adaption}. 
The reduced EQANIS model is based on the expression~\eqref{eq:sysFunc2DCheb}, but instead of utilizing the full series of Chebyshev polynomials in $P^{(2)}_k$, it retains only a single summand at index $\lambda_k^*$, which determines the order of the Chebyshev polynomials used in each frequency component. 
The validity of this reduction depends on the type of excitation. It has been proposed specifically for Lissajous-type excitation patterns, but its applicability to other FFP trajectories remains unexplored.
As we stick to a two-dimensional excitation and reconstruction later on, we limit the description in the following to this case. Therefore, we only consider two components for all equations, whereas we previously considered three. 
That said, for an excitation of the form 
\begin{equation*}
\HD(t) = \begin{pmatrix} 
A_x\sin(2\pi f_x t +\varphi_x) \\ 
A_y\sin(2\pi f_y t +\varphi_y)
\end{pmatrix}
\end{equation*}
with ${A_x,A_y\in\R\backslash\{0\}}$,   ${\varphi_x,\varphi_y}\in \R$, and frequencies defined as ${f_x=\frac{\fB}{\NB+1}}$, ${f_y=\frac{\fB}{\NB}}$, where ${\fB>0}$, $\lambda_k^*$ is proposed as 
\begin{equation*}
    \lambda_k^* =  \operatorname{round}\!\left(\frac{2\NB k}{2\NB^2+2\NB+1}\right).
\end{equation*}
For other cases, including three-dimensional excitation, we refer to~\cite{droigk2022direct}.

With the restriction to one single summand, the system function in the Fourier domain can be approximated by

\begin{equation}
	\begin{aligned}
		\bm{s}_{ k}(\bm{x}) &\approx \bm{M}_{ k}\!\!\int_{\mathbb R^2} \left[\frac{\partial^2}{\partial z_1 \partial z_2}\mathscrbf{E}(\beta \bm{G} \bm z; \mathbb O(\bm x))\right]_{\scriptscriptstyle \bm z = \bm x- \bm y}\\
        &\hspace{4.5cm} \times S^{(2)}_k(\bm{y}, \lambda^*_k) \dt[\bm{y}]
	\end{aligned}
	\label{eq:sysFuncCompAnisoReduced}
\end{equation}
with
\begin{equation}
    S_k^{(2)}(\bm x, \lambda^*_k) = \frac{\sgn(\tfrac{A_x A_y}{G_x G_y})\iUnit^{\lambda_k^*}\eUnit^{\iUnit\theta_k^*}}{\pi^2}  
    V_{n^*_k}\!\left(\frac{G_x}{A_x}x_1\right)V_{m_k^*}\!\left(\frac{G_y}{A_y}x_2\right)
\end{equation}
and
\begin{equation}
    V_n(\xi) = 
    \begin{cases} 
    \rect\!\left( \frac{\xi}{2}\right)\left(-\frac{U_{|n|-1}(\xi)\sqrt{1-\xi^2}}{|n|}\right),& \text{if } |n|>0 \\
    \frac{\pi}{2}\sgn(\xi+1) -\rect\!\left( \frac{\xi}{2}\right)\arccos(\xi),& \text{if } |n|=0,
    \end{cases}
\end{equation}
where $U_m:\R\to\R$ denotes the Chebyshev polynomials of second kind with order $m\in\N_0$. The orders $n_k^*$ and $m_k^*$ of the Chebyshev polynomials depend on $\lambda_k^*$ and follow as ${n_k^* =-k + \lambda_k^* (\NB+1)}$ and ${m_k^* = k - \lambda_k^*\NB}$.

Equation~\eqref{eq:sysFuncCompAnisoReduced} shows that the system function component can be expressed as a convolution between the tensor-product of Chebyshev polynomials with a spatially varying convolution kernel 
\begin{equation}
\bm K(\bm x,\bm y) = \tfrac{\partial^2}{\partial y_1 \partial y_2}\mathscrbf{E}(\beta \bm{G} \bm y; \mathbb O(\bm x)).    
\end{equation}

\subsection{Generalized DCR with EQANIS}

By restricting the infinite series to a single summand, as shown in~\eqref{eq:sysFuncCompAnisoReduced}, the convolved particle distribution can be reconstructed as a weighted sum of tensor products of Chebyshev polynomials of the second kind. To this end, the convolution with the kernel $\bm K(\bm x, \bm y)$ is applied directly to the particle distribution. Consequently, the frequency components of the measured voltage signal correspond to the coefficients of a Chebyshev series expansion of the convolved particle distribution. 
Let $\tilde{\bm K}(\bm x,\bm y) = {\bm K}(\bm x,-\bm y)$. Then, for a two-dimensional consideration, the convolved particle distribution is given by
\begin{align}
\begin{split}
\tilde{\bm c} (\bm x) = & \int_{\mathbb{R}^2} c(\bm z) \bm K( \bm x, \bm z- \bm G^{-1} \bm A \bm x) \dt[\bm{z}]\\
= &\Bigl(c(\bm y) \ast \bm{K}(\bm x,-\bm y)\Bigr)\left(\bm G^{-1} \bm A \bm x\right)\\
= &\left(c(\bm y)\ast \tilde{\bm{K}}(\bm x,\bm y)\right)\left(\bm G^{-1} \bm A \bm x\right)
\label{eq:ctilde_c_relation}
\end{split}
\end{align}
with $\bm A = \operatorname{diag}(A_x, A_y) $. The convolution is performed component-wise and with respect to the second variable of $\bm K(\bm x,\bm y)$ and $\tilde{\bm K}(\bm x,\bm y)$. Note that the last bracket does not indicate a multiplication, but specifies the argument at which the convolution result is evaluated.
This convolved SPIO distribution can be approximated using the following relationship~\cite{droigk2023adaption}:
\begin{equation}
\tilde{\bm c}(\bm x) \approx \sum_{k\in\mathbb{K}} \frac{4 n_k^* m_k^*}{ \det\!{({\bm{G}^{-1}\bm{A}})}} \bm{M}_k^{-1} \bm u_k \iUnit^{\lambda_k^*} U_{|n_k^*|-1}(x_1) U_{|m_k^*|-1}(x_2).
\label{eq:recoctilde}
\end{equation}
The set $\mathbb{K}$ contains the frequency components used for reconstruction. However, there are limitations on its selection. On the one hand, it must include only frequency components where $n_k^*>0$ and $m_k^*>0$. On the other hand, it must not include frequency components leading to duplicate values of $|n_k^*|$ and $|m_k^*|$. If two frequency components correspond to the same values of  $|n_k^*|$ and $|m_k^*|$, the larger one should be preferred, as it has a higher energy level and is therefore less affected by noise.

After reconstructing $\tilde{\bm c}$, a deconvolution step is required to achieve a high-quality result. 
In the original DCR method presented in~\cite{droigk2022direct}, which is based on the equilibrium model without anisotropy, a conventional deconvolution is sufficient. This is possible because in that case the convolution kernel $\bm K(\bm x,\bm y)$ is independent of $\bm x$ and corresponds to the second partial derivative of the multidimensional Langevin function. However, when anisotropy is incorporated in the equilibrium model, the conventional convolution transforms into a spatially varying convolution with the kernel $\tilde{\bm K}(\bm x,\bm y)$.

\begin{figure}
\begin{center}
 \includegraphics[width=0.45\textwidth]{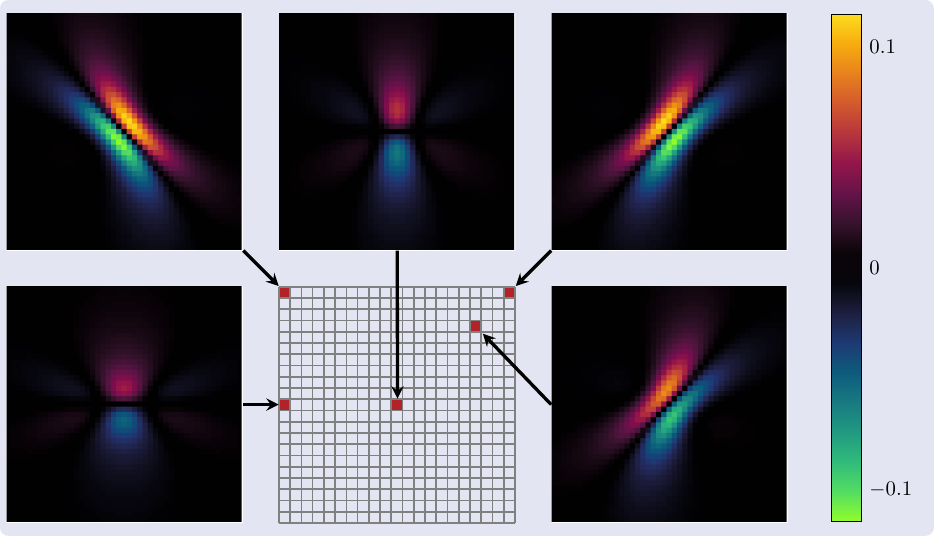}
\end{center}
\caption{Example of the anisotropy kernel $\bm K (\bm x, \bm y) = \tfrac{\partial^2}{\partial y_1 \partial y_2}\mathscrbf{E}(\beta \bm{G} \bm y; \mathbb O(\bm x))$ at different positions of the discrete drive-field field-of-view, which is shown in the center of the bottom row.
The shape of the convolution kernels changes depending on the position $x$ in the field-of-view. They exhibit a certain symmetry, but the corresponding symmetry axis also changes its orientation. The maximum amplitude also changes depending on the location. The variable $y$ refers to the spatial coordinates within the kernel.}
\label{fig:anisotropykernel}
\end{figure}

A visualization of this convolution kernel at different positions within the FOV is shown in Fig.~\ref{fig:anisotropykernel}. At the center of the FOV, the kernel is equal to that of the DCR-EQ method,  where it can be treated as a standard convolution across the entire FOV. 
However, towards the edges and corners of the FOV, the kernel shape undergoes significant changes. These variations are primarily driven by the preferred direction of the anisotropy and the anisotropy strength, which increases with distance from the center. 

Due to the spatial variation of the kernel, the deconvolution process must be adapted accordingly. For this purpose, a convolution matrix $\bm M$ was introduced in~\cite{droigk2023adaption}. Let $N\in\mathbb{N}$ with $N=N_1 N_2$ denote the number of pixels in the FOV and $M\in\mathbb{N}$ with $M=M_1M_2$ the number of pixels at which the kernel is evaluated. The flexibility of the DCR method allows $N$ and $M$ to be freely chosen. Consequently, the convolution matrix is of size ${\bm M\in\mathbb{R}^{2N \times N+M+1}}$.
Since $\tilde{\bm c}$ is reconstructed separately for each receive path, and the kernel $\bm K$ is also path-specific (though symmetric), the matrix $\bm M$ is composed of two submatrices, ${\bm M_x, \bm M_y \in\mathbb{R}^{N\times N+M+1}}$, which describe the spatially variant convolution for the $x$- and $y$-receive path, respectively. 
The deconvolution problem can then be formulated as the following optimization problem:
\begin{equation}
\min_{\bm c_d } \quad
\left\|
\begin{pmatrix}
\bm M_x\\\bm  M_y
\end{pmatrix} 
\bm c_d 
- 
\begin{pmatrix}
\tilde {\bm c}_{1,d}\\ \tilde{\bm c}_{2,d}
\end{pmatrix} 
\right\|_2^2
+ \lambda \mathcal R(\bm c_d).
\end{equation}
Here, $\tilde{\bm c}_{1,d}$ and $\tilde{\bm c}_{2,d}$ represent the discrete first and second components of the reconstructed $\tilde{\bm c}$ from~\eqref{eq:recoctilde}. The parameter ${\lambda\in\mathbb{R}^+}$ controls the influence of the regularization function ${\mathcal R:\mathbb{R}^{N+M+1}\rightarrow \mathbb{R}}$. 
This minimization problem can be solved iteratively. However, if the regularization function is chosen as $\mathcal R(\cdot)=\|\cdot\|_2^2$, an analytical solution is also possible. 
The obtained $\bm c_d$ now has the size $N+M+1$  and is therefore larger than the DF-FOV.  This is useful to reduce artifacts, but no information outside the DF-FOV has been transferred by $\tilde{\bm c}$.  Therefore, it makes sense to crop $\bm c_d$ back to the size $N=N_1\cdot N_2$ of the DF-FOV. 


\subsection{Approximating the Operator}

The evaluation of the operator in \eqref{eq:ctilde_c_relation} can be accelerated by introducing a rank-$p$ approximation of the spatially variant kernel $\bm K(\bm x, \bm y)$, allowing the variables $\bm x$ and $\bm y$ to be separated. This is also true for $\tilde{\bm K}(\bm x, \bm y)$. 
Although the approximation is possible for three-dimensional spatial variables, for the sake of uniformity, we will restrict ourselves to a two-dimensional approach, as was done in the previous subsections. Then, the separation of the kernel reads
\begin{equation} \label{eq:kernel_approximation}
    \bm K(\bm x, \bm y) \approx \sum_{i=1}^{p} \bm b_{i}(\bm y)  \odot \bm w_{i}(\bm x),
\end{equation}
where $\odot$ denotes the Hadamard product and ${\bm b_{i}, \bm w_{i}: \mathbb{R}^2\to \mathbb{R}^2}$ build up the separation. 
Besides, the flipped version of $\bm K(\bm x, \bm y)$ can be decomposed as
\begin{equation} \label{eq:kernel_approximation_flipped}
    \tilde{\bm K}(\bm x, \bm y) \approx \sum_{i=1}^{p} \tilde{\bm b}_{i}(\bm y)  \odot \bm w_{i}(\bm x),
\end{equation}
with $\tilde{\bm b}_{i}(\bm y) ={\bm b}_{i}(-\bm y)$.

Among all possible functional decompositions of this form, the Karhunen-Loève decomposition minimizes the total mean square error \cite{Wang2008}. In a discretized setting, \eqref{eq:kernel_approximation} can be computed using principal component analysis, where the $p$ components corresponding to the largest eigenvalues are selected. Substituting \eqref{eq:kernel_approximation} into \eqref{eq:ctilde_c_relation}, we obtain the following approximation of the operator:
\begin{equation}
	\begin{aligned}
		\tilde{\bm c} (\bm x) & \approx \sum_{i=1}^{p} \int_{\mathbb{R}^{2}} c(\bm z) \bm b_{i}(\bm z - \bm G^{-1} \bm A \bm x)\odot \bm w_{i}(\bm x)\dt[\bm z]
        \\
         &= \sum_{i=1}^{p} \bm w_{i}( \bm x) \odot  \biggl(c(\bm y) * \tilde{\bm b}_{i}(\bm y)\biggr)(\bm G^{-1} \bm A \bm x).
	\end{aligned}
	\label{eq:approximated_fwd_operator}
\end{equation}
This shows that the operator can be approximated as a weighted sum of convolutions. For the operator in~\eqref{eq:ctilde_c_relation}, we can formulate its adjoint operator as
\begin{equation}
  \bm1 ^\intercal \int_{\mathbb{R}^{2}} \tilde{\bm c}(\bm z)\odot \bm K(\bm z, \bm y-\bm G^{-1} \bm A \bm z)\dt[\bm z],
\end{equation}
where $\bm 1 = (1,1)^\intercal$ denotes the vector containing ones. 
We can approximate this expression using the same rank-$p$ decomposition of the kernel, yielding
\begin{equation} \label{eq:approximated_adj_operator}
\begin{aligned}
   &\bm1 ^\intercal  \sum_{i=1}^{p} \int_{\mathbb{R}^{2}} \tilde{\bm c}(\bm z)\odot \bm w_{i}(\bm z) \odot \bm b_{i}( \bm y-\bm G^{-1} \bm A \bm z)\dt[\bm z]\\
    =&\bm1 ^\intercal  \sum_{i=1}^{p} \left(\left(\bm w_{i}(\bm x) \odot  \tilde{\bm c}(\bm x) \right)\ast \bm b_{i}(\bm G^{-1} \bm A \bm x) \right)(\bm A ^{-1} \bm G \bm y).
\end{aligned}
\end{equation}
 Since the decomposition can be precomputed before reconstruction, it provides a computational advantage by enabling efficient evaluation of both the forward and adjoint operators.

The evaluation of these approximations enables fast implementations of the convolutions, such as using the fast Fourier transform. This reduces the computational complexity from $\mathcal{O}(N^2)$ required for direct evaluation of \eqref{eq:ctilde_c_relation}, to $\mathcal{O}(N\log N)$.

For a two-dimensional kernel represented on a grid of size $N_1\times N_2 \times M_1 \times M_2$, $NM$ elements have to be stored. In contrast, an approximation following the form of~\eqref{eq:kernel_approximation} in the same grid reduces the storage requirement to only $p(N+M)$ elements. Furthermore, since the summands in~\eqref{eq:approximated_fwd_operator} and~\eqref{eq:approximated_adj_operator} are independent, the computations can be parallelized.


This approach has been successfully applied to a similar problem in optics~\cite{Lauer2002}, and preliminary reconstruction results for MPI were presented in \cite{Duran2025}.

\section{Materials}

To assess the reconstruction quality of the rank-$p$ approximation in the spatially variant deconvolution of the DCR-EQANIS for different values of $p$, and to compare it with both the original operator and the deconvolution of the DCR-EQ, simulations and measurements were conducted. The same scanner parameters were employed in the simulations as in the real measurements.  

The experiments utilized a two-dimensional imaging sequence in the $xy$-plane of the scanner (horizontal orientation). A gradient of $-1 ~\text{T}/\text{m}/\mu_0$ was applied along the $x$- and $y$-axes, while a gradient of $2 ~\text{T}/\text{m}/\mu_0$ was applied along the $z$-axis. The drive field frequencies were set to $f_x = 2.5~\text{MHz} / 102$ and $f_y = 2.5~\text{MHz} / 96$, with drive field amplitudes of $A_x, A_y = 12~\text{mT}/\mu_0$. This configuration resulted in a Lissajous sampling trajectory with a frequency ratio of $\tfrac{f_y}{f_x}=\frac{17}{16}$, i.e., $\NB =16$.

\subsection{Approximation quality}
To evaluate the approximation of the spatially varying kernel using the rank-$p$ approximation, the mean squared error (MSE) between the convolution kernels derived from the model and their approximated counterparts was computed. 
For this purpose, a drive-field FOV  (DF-FOV) of $21 \times 21$ pixels was assumed, and the corresponding convolution kernel was determined at each position. The size of the convolution kernels was chosen to be larger than the DF-FOV, measuring $41\times41$ pixels. The MSE was computed in two ways: first, as an overall metric across the entire DF-FOV for ranks ranging from $p=1$ to $p=60$; and second, as a position-dependent measure, evaluated individually at each location within the DF-FOV.

\subsection{Simulations}

To obtain a simulated voltage signal, a system matrix was simulated based on the equilibrium model with anisotropy as described in~\cite{maass2024equilibrium}. The system matrix was simulated within a FOV of size $\SI{34}{\mm} \times \SI{34}{\mm}$ on an equidistant grid of size $201 \times 201$ pixels. The particle core size was set to $\SI{21}{\nano\meter}$ and an anisotropy strength $\ak$ with $\Kanismax = \SI{2000}{\joule\meter}^{-3}$ was used. This choice was based on the measurement data, where this combination of parameters was found to be suitable.

Frequencies below \SI{80}{\kHz} and above \SI{450}{\kHz} were discarded. The resulting system matrix was then used to generate six different voltage signals by multiplying it with the vectors of six different phantoms. No noise is added to better observe the isolated effects of the operator approximation. 

The DCR was applied by computing \eqref{eq:recoctilde} on a grid of size $21\times21$ pixels. The deconvolution of the obtained $\tilde{\bm c}$ was then performed using three approaches: (i) the Langevin kernel corresponding to DCR-EQ, (ii) the spatially varying anisotropy kernel corresponding to DCR-EQANIS using a convolution matrix, and (iii) the spatially varying anisotropy kernel corresponding to DCR-EQANIS using the rank-$p$ approximation of the operator for different values of $p$.

Before deconvolution, $\tilde{\bm{c}}$ was padded using repetitive boundary conditions with a width of 2 pixels in each spatial direction. For all methods, optimization was performed using the fast iterative shrinkage-thresholding algorithm (FISTA) \cite{Beck2009} with Tikhonov regularization. For all phantoms, the same regularization parameter was used. It was optimized for each method by visual inspection. However, a regularization parameter of $\lambda = 10^{-1}$ has proven to be well-suited for all methods. 

\subsection{Measurements}

The preclinical MPI scanner from Bruker (Ettlingen, Germany) was used to acquire the experimental data. This dataset, measured using fluid Perimag at a concentration of $\SI{10}{\mg_{\text{Fe}}\per\ml}$, is also referenced in~\cite{maass2024equilibrium}. The system matrix was obtained by shifting the $\Delta$-sample to $17 \times 15$ positions, covering a FOV of $\SI{34}{\milli\meter} \times \SI{30}{\milli\meter}$.

In addition to the system matrix, six different phantoms were measured to investigate the impact of modeling errors on image reconstruction. The snake phantom consisted of five cubic rods with a cross-section of $\SI{2.5}{\mm} \times \SI{2.5}{\mm}$ and lengths of \SI{20}{\mm}, \SI{17.5}{\mm}, \SI{15}{\mm}, \SI{8.75}{\mm}, and \SI{5}{\mm}, arranged in a snake-like pattern. Three resolution phantoms were designed with two rods of lengths \SI{20}{\mm} and \SI{17.5}{\mm}, positioned in parallel at distances of \SI{3}{\mm}, \SI{5}{\mm}, and \SI{7}{\mm}. The fifth phantom featured a shape resembling an ice cream cone, with a conical lower part and a spherical upper part. The sixth phantom consisted of the $\Delta$-sample, positioned \SI{6}{\mm} away from the center along both the $x$- and $y$-axes. For all measurements, background correction was applied using the method described in~\cite{knopp2019correction}.

The original image was reconstructed from the resulting voltage signal using various methods. For all methods, the drive-field FOV (DF-FOV) was reconstructed on a grid of size $21\times21$.

For comparison, image reconstruction was performed using both measured and simulated system matrices. Since the measured system matrix was only available for an FOV size of $17\times 15$, the corresponding reconstruction results were linearly interpolated and subsequently cropped to match the DF-FOV.

In addition, the simulated system matrices corresponding to the equilibrium model, with and without anisotropy, were used for reconstruction. The simulation parameters were chosen based on~\cite{maass2024equilibrium}, where they yielded the best results for the same measurement data. The weighted and regularized least-squares problem  
\begin{equation}
   \underset{\bm{c}\geq \bm{0}}{\text{min}} \left(\Vert \bm{W} ( \bm{S} \bm{c} - \bm{u}) \Vert_2^2 + \lambda \Vert \bm{c} \Vert_2^2 \right)
\end{equation}
was solved, where the matrix $\bm W$ was chosen such that the noise was whitened according to a diagonal covariance matrix. The regularization parameter was optimized by visual inspection and was chosen as $\lambda = 10^{-0.2}$ for both the equilibrium model with anisotropy and the model without anisotropy. For the reconstruction with the measured system matrix, the regularization parameter was optimal at $\lambda = 10^{-1}$. 

Furthermore, DCR was applied to obtain $\tilde{\bm{c}}$. As in the simulations, $\tilde{\bm{c}}$ was padded using repetitive boundary conditions with a width of $2$ pixels in each spatial direction before deconvolution.  

For the deconvolution step, the equilibrium model without anisotropy was used first, corresponding to the DCR-EQ. This approach employs the second partial derivative of the Langevin function as the kernel for deconvolution. Furthermore, DCR-EQANIS adapted to the anisotropy model in~\cite{droigk2023adaption}, was applied. In this case, the convolution matrix was constructed to perform deconvolution with the spatially varying kernel.  

Finally, the method presented in this work, which enables a fast approximation of the operator, was used to perform deconvolution within the DCR-EQANIS framework. Different numbers of basis functions were tested to approximate the original kernel.  

In order to determine the model parameters, a parameter search for the particle core size and the anisotropy strength was carried out. The best optical results were achieved with a particle size of $\SI{21}{\nano\meter}$ and an anisotropy constant of ${\Kanismax = \SI{2000}{\joule\per\cubic\meter}}$. 

For all deconvolution methods, Tikhonov regularization was applied with an optimized regularization parameter. As in the simulation experiments, a regularization parameter of ${\lambda=10^{-1}}$ was ideal for all methods. 

\subsection{Runtime and memory consumption}

In addition to evaluating reconstruction quality, the runtime and memory consumption of the different methods were also analyzed. For this purpose, the grid size of the drive-field FOV was varied among $11\times11$, $21\times21$, ..., $61\times 61$ pixels. 
Since FISTA was used for optimization in all methods, the mean time per iteration of FISTA was measured for runtime comparison. To assess memory consumption, the required inputs for FISTA in the two methods were compared. Specifically, this included the memory required to store the convolution matrix and the memory needed to represent the convolution kernels using a rank-$p$ approximation. 

\section{Results}
In the following, the results of the respective experiments are presented and described.
\subsection{Approximation quality}

\begin{figure}
 \begin{center}
    \includegraphics[width=0.49\textwidth]{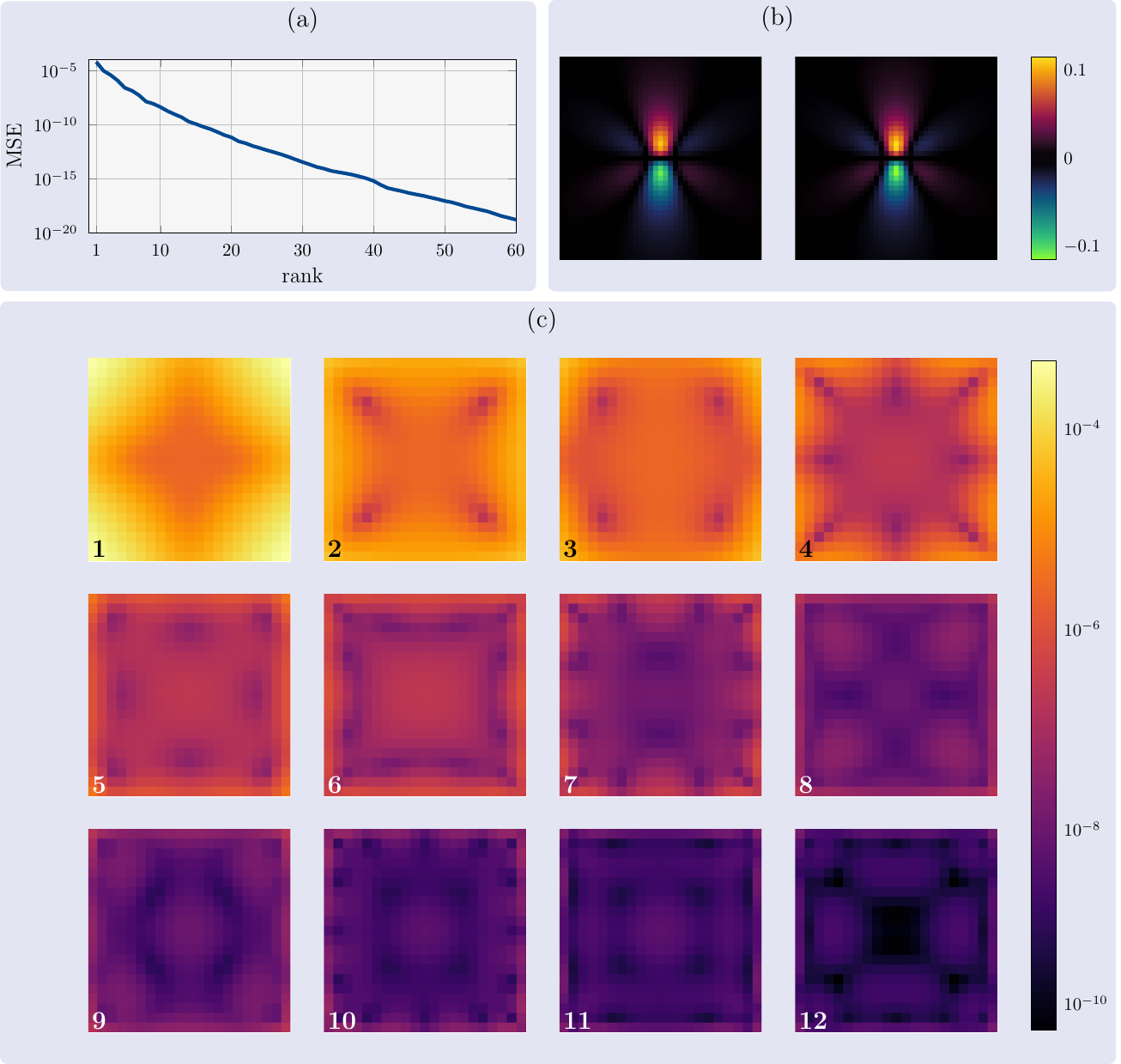}
 \end{center}
 \caption{Behavior of the mean squared error (MSE) of the rank-$p$ approximation of the spatially varying convolution kernel and comparison of the first principal component. (a) rank vs MSE. The error is averaged over all positions within the convolution kernels and all positions of the FOV.  (b) Comparison of the Langevin kernel (left) to the first principal component (right) of the rank-$p$ approximation. (c) MSE distribution inside the field-of-view for ranks $p=1,2,...,12$. The error is averaged over all positions within the convolution kernels. The rank is given in the lower left corner of the corresponding MSE heatmap.}
 \label{fig:MSEDistribution}
 \end{figure}

The distribution of the MSE, which quantifies the accuracy of the rank-$p$ approximation,  is shown in Fig.~\ref{fig:MSEDistribution}a. The MSE decreases exponentially with increasing rank. 

In Fig.~\ref{fig:MSEDistribution}b, the first principal component of the decomposition is shown and compared to the convolution kernel of the equilibrium model without anisotropy, i.e. the second partial derivative of the Langevin function. Although the first basis function is not entirely identical, it exhibits a very high degree of similarity. This indicates that deconvolution using the rank-1 approximation closely resembles the classical DCR-EQ approach. However, the additional weighting of this component enables adaption to the varying amplitudes of the convolution kernel at different positions within the FOV - see Fig.~\ref{fig:anisotropykernel}.

The spatial distribution of the MSE is illustrated in Fig.~\ref{fig:MSEDistribution}c.  At low ranks, the error is particularly large at the edges of the FOV. The deviation of the convolution kernel from the first principal component is most significant in the corners, which is reflected in the MSE distribution. However, as the rank increases, the error decreases across all positions within the FOV, ultimately converging towards zero.

\subsection{Simulations}

\begin{figure}
 \begin{center}
    \includegraphics[width=0.49\textwidth]{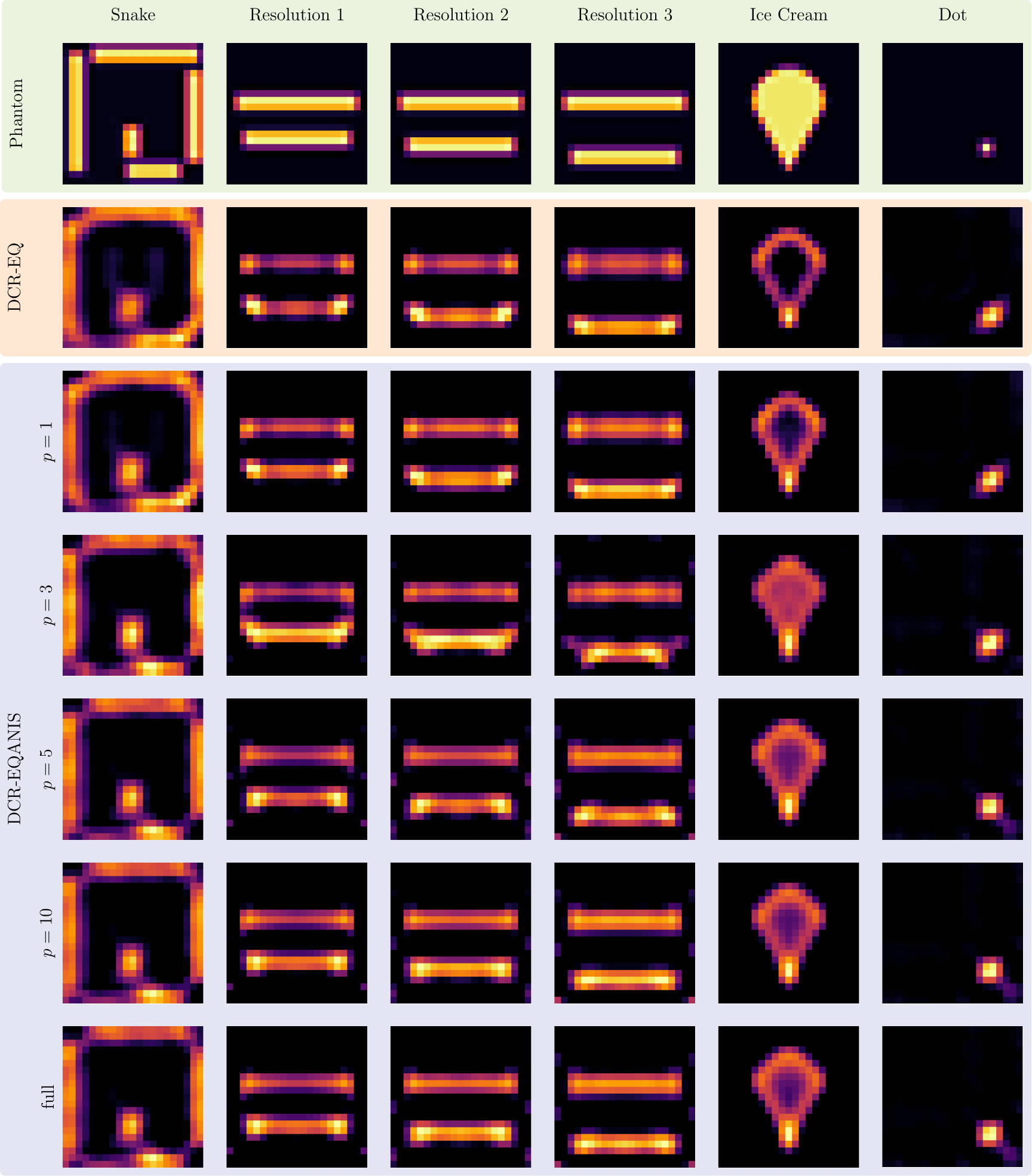}
 \end{center}
 \caption{Reconstruction results of the  simulation experiments with same computation of $\tilde{\bm c}$ for all methods, followed by different deconvolution methods. 1st row: phantoms used for signal generation. 2nd row: deconvolution results of the DCR-EQ. 3rd - 7th row: deconvolution results for rank-$p$ approximation of DCR-EQANIS with $p=1,3,5,10$ and full rank.}
 \label{fig:simulations}
 \end{figure}

Fig.~\ref{fig:simulations} presents the reconstruction results of the simulation experiments using six different phantoms. 

A comparison between DCR-EQANIS with full rank and DCR-EQ reveals that the former better preserves the original shape of the phantoms. This improvement is evident for all phantoms except the ice-cream phantom, where the reconstructed shapes are highly similar across all methods.

For the resolution phantoms, the individual beams exhibit slight curvature at the ends when using DCR-EQ, whereas DCR-EQANIS maintains their straight shape. Similarly, the dot phantom is better restored by DCR-EQANIS, as it appears to be rather elliptical than circular in the DCR-EQ reconstruction.

In the DCR-EQ reconstruction, the snake phantom appears to have rounded elements in the corners of the image that merge into each other, while the DCR-EQANIS preserves the angular shape and successfully reconstructs the small gaps between the bars. 

The similarity of the ice-cream phantom reconstructions across methods is likely because this phantom does not extend into the edge regions, particularly the corners. As shown in Fig.~\ref{fig:anisotropykernel}, the convolution kernels exhibit only minor variations in the inner areas, which explains the consistent reconstruction quality.  

Regarding the different rank-$p$ approximations of DCR-EQANIS, it is noteworthy that for $p=1$, the results closely resemble those of DCR-EQ. This is expected due to the similarity of the convolution kernel, as illustrated in Fig.~\ref{fig:MSEDistribution}. With $p=5$, the reconstruction is already highly similar to the full-rank solution, with only minor deviations visible in background artifacts. Additionally, the bars of the resolution phantoms appear slightly less uniform compared to higher-rank reconstructions. At $p=10$, no perceptible difference can be observed compared to the full-rank reconstruction.

At this point, a brief explanation is warranted regarding the ``hole'' observed in the ice-cream phantom reconstructions, which is particularly pronounced in the DCR-EQ reconstruction. 
This phenomenon arises due to the realistic frequency selection used in these experiments, where frequencies below \SI{80}{\kHz} were excluded from the reconstruction. However, the discarded frequency components predominantly contain low-frequency information and are represented by tensor products of low-order Chebyshev polynomials. Omitting these components can thus lead to the appearance of holes within large homogeneous areas. A test reconstruction not shown here that included the low-frequency components confirmed this effect, as the hole in the ice-cream phantom disappeared. 

\subsection{Measurements}
\begin{figure}
 \begin{center}
    \includegraphics[width=0.49\textwidth]{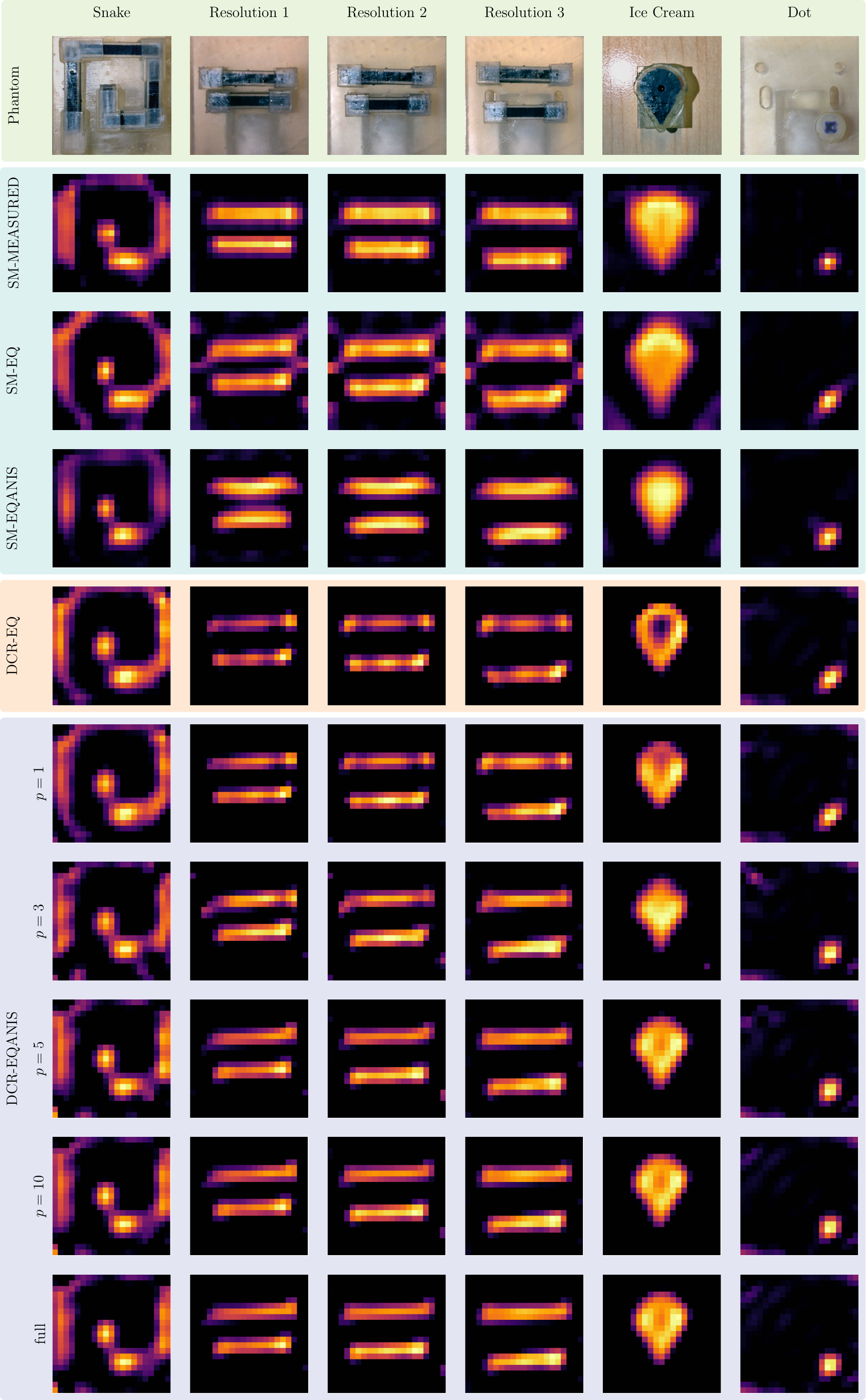}
 \end{center}
 \caption{Reconstruction results of the measurement experiments.
 1st row: Photos of the phantoms. 
 2nd row: Reconstruction using the measured system matrix (SM-MEASURED). 
3rd row: Reconstruction using a simulated system matrix following the equilibrium model without anisotropy (SM-EQ). 4th row: Reconstruction using a simulated system matrix following the equilibrium model with anisotropy (SM-EQANIS). 5th row: Reconstruction using the direct Chebyshev reconstruction with the Langevin kernel used for deconvolution (DCR-EQ). 6th-10th row:  Reconstruction using the direct Chebyshev reconstruction with the spatially varying anisotropy kernel used for deconvolution (DCR-EQANIS) with rank-$p$ approximation for $p=1,3,5,10$ and full rank.}
 \label{fig:measurements}
 \end{figure}

When reconstructing the measurement data, similar observations can be made as in the simulation results regarding the preservation of the phantom shapes, as shown in Fig.~\ref{fig:measurements}. Here too, the bars of the resolution phantoms appear less uniform in the DCR-EQ reconstruction compared to the full-rank DCR-EQANIS, the point phantom exhibits slight distortion, and the hole within the ice-cream phantom, as discussed previously, is more pronounced. As the rank of the rank-$p$ DCR-EQANIS increases, the reconstruction progressively converges towards the full-rank DCR-EQANIS result. The comparison between rank-$1$ DCR-EQANIS and DCR-EQ reveals slightly more pronounced differences in the resolution phantoms and the ice-cream phantom than in the simulation. While the convolution kernels are very similar, the additional position-dependent weighting within the FOV enhances image quality and also appears to partially mitigate the hole in the ice-cream phantom.

System matrix reconstructions were also utilized as additional comparison methods for the measurement data. When comparing the SM-EQ reconstruction with all other methods, it exhibits particularly pronounced background artifacts. Although DCR-EQ is based on the same model and initially reduces these artifacts further, they are almost entirely suppressed in the latter. Background artifacts are only weakly visible in the point phantom but remain significantly less pronounced compared to the SM-EQ reconstruction.

The comparison of the DCR-EQANIS with the SM-EQANIS yields interesting insights. Contrary to the expectations expressed in~\cite{droigk2022direct}, which suggested that no improvement in image quality could be expected over a simulated system matrix reconstruction due to the additional simplifications introduced for the DCR, the results indicate the opposite.

While the upper bar of the snake phantom may be better reproduced with the SM-EQANIS reconstruction, DCR-EQANIS surpasses its image quality in other regions and across the other phantoms. The bars of the resolution phantoms appear sharper and more angular, without the smeared transitions observed in the resolution-$1$ phantom. Additionally, the boundaries of the ice-cream phantom are more distinct, and the point phantom appears rounder and more uniformly shaped. These improvements are likely attributable to differences in the optimization problem, allowing for a lower level of regularization in DCR-EQANIS without introducing significant background artifacts.  

A comparison between DCR-EQANIS and the state-of-the-art reconstruction using a measured system matrix (SM-MEASURED) reveals largely comparable results across all phantoms. The overall shapes of the snake, dot, ice-cream, and resolution phantoms are reconstructed with high fidelity by both methods. Minor background artifacts are observed in the DCR-EQANIS reconstruction of the snake and dot phantoms, which are absent in the SM-MEASURED results. However, DCR-EQANIS produces noticeably sharper and less blurred reconstructions for the remaining phantoms, highlighting its ability to preserve structural details despite the absence of a measured system matrix.

\subsection{Runtime and memory consumption}
\begin{figure}
    \begin{center}
        \includegraphics[width=\columnwidth]{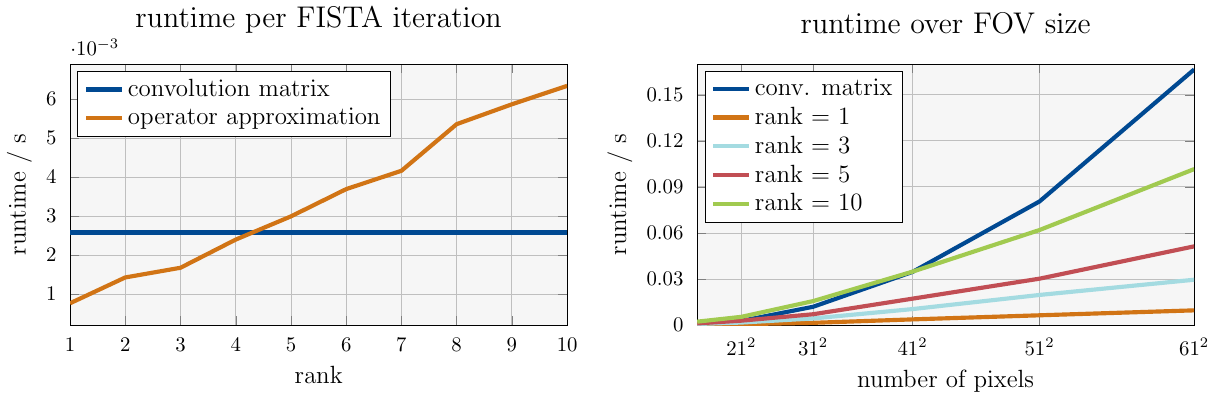}
    \end{center}
    \caption{Runtime comparison of the DCR-EQANIS with, on the one hand, deconvolution via convolution matrix and, on the other hand, the proposed method via rank-$p$ operator approximation. The average time required for a single FISTA iteration is shown in each case. Left: Behavior of the runtime as a function of the rank for a fixed field-of-view size  $N=21^2 = 441$. Right: Runtime as a function of the FOV size. }
    \label{fig:runtime}
\end{figure}

The results of the runtime comparison in Fig.~\ref{fig:runtime} indicate that the runtime of the DCR-EQANIS deconvolution with \mbox{rank-$p$} approximation  increases linearly with the rank size. This trend is evident in the left graph, where the average runtime per FISTA iteration is plotted as a function of the rank for a fixed number of pixels, $N=21^2$, as used in the previous experiments. For this configuration, the rank-$p$ approximation with $p<5$ is faster than the computation using a convolution matrix within FISTA. 
As illustrated in the right graph of Fig.~\ref{fig:runtime}, this distinction shifts depending on the number of pixels in the FOV. Since the calculation time for deconvolution using a convolution matrix increases quadratically, whereas the rank-$p$ approximation scales with $N \log(N)$, the rank-10 approximation is already computed at a comparable speed for an FOV of size $41 \times 41$ ($N=1681$), and becomes even faster as the FOV resolution increases. When using a rank of $p=5$, which in previous experiments produced results nearly indistinguishable from those obtained with full-rank computations, the reconstruction is already twice as fast as the convolution-matrix based solution for an FOV with a grid size of $41 \times 41$. As the FOV resolution increases further, the computational advantage of the rank-$p$ approximation continues to grow. 

\begin{figure}
    \begin{center}
        \includegraphics[width=\columnwidth]{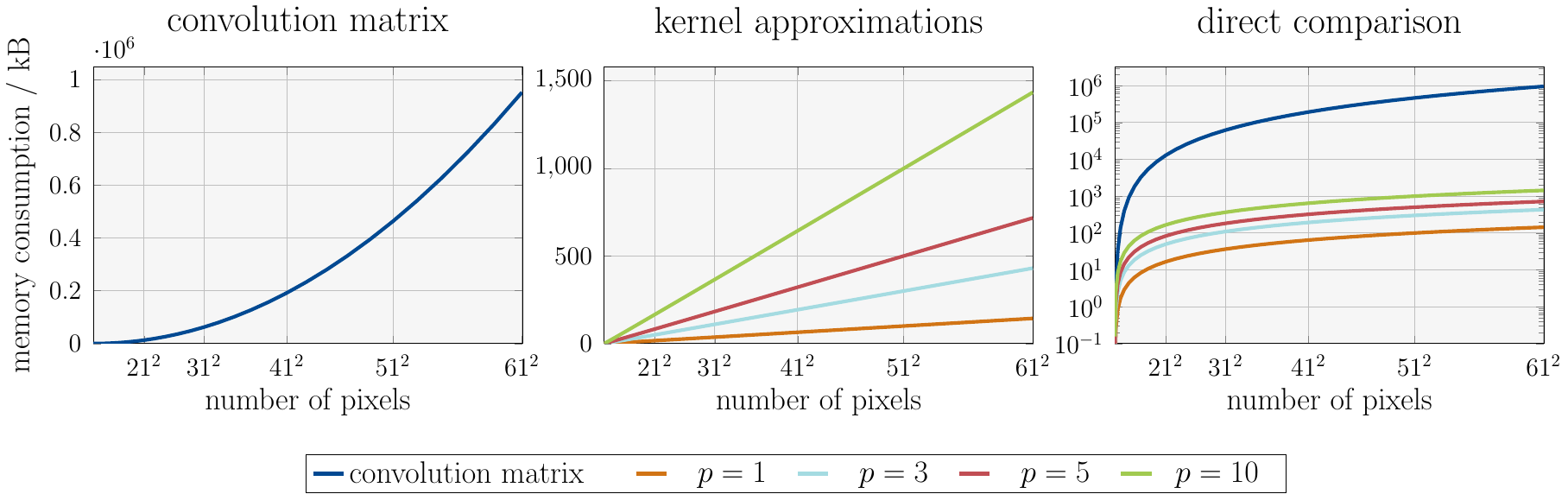}
    \end{center}
    \caption{Memory consumption of the convolution matrix and the rank-$p$ approximation for different $p$. Left:  Memory consumption of the convolution matrix. Middle: Memory consumption of the rank-$p$ approximation for $p=1,3,5,10$. 
Right: Direct comparison of the approaches.}
    \label{fig:memory}
\end{figure}

Finally, the memory consumption of the two methods was compared, with the results presented in Fig.~\ref{fig:memory}. When a convolution matrix is set up to perform the deconvolution inside the DCR-EQANIS, its memory consumption scales with $\mathcal{O}(N^2 + NM)$. This quadratic growth with respect to the number of pixels $N$ is evident in the curve on the left.
With the rank-$p$ approximation and the associated fast operator, the convolution matrix is no longer required. Instead, only a representation of the various convolution kernels is needed, which can be stored within $\mathcal{O}(p(N+M))$. The exact behavior is shown in the middle graph for FOVs of different sizes. The linear progression of the memory consumption is evident. Even with the relatively small number of pixels used here ($N=441$), the difference in memory consumption is significant: the convolution matrix requires \SI{12.5}{\mega\byte}, whereas the rank-$p$ approximation with $p=10$ requires only \SI{165}{\kilo\byte}, with lower ranks demanding even less. For an FOV of size ${N=51 \cdot 51 = 2601}$, the convolution matrix already requires \SI{452}{\mega\byte}, while the rank-$p$ approximation with $p=10$ remains below \SI{1}{\mega\byte}. This increasing difference is also clearly visible in the graph on the right. 

\section{Discussion}

It has already been demonstrated in~\cite{maass2024equilibrium} that extending the equilibrium model to incorporate anisotropy effects provides qualitative advantages in the reconstruction of measurement data when using a system matrix for model-based reconstruction.
The present study now shows that this advantage also holds for model-based Chebyshev reconstruction, which relies on the respective models in a reduced form.
Furthermore, the quality of the DCR reconstruction was superior to that of model-based system matrix reconstructions for most phantoms and comparable to state-of-the-art reconstructions using a measured system matrix. However, a key limitation of the DCR remains its restriction to reconstructing only within the DF-FOV, since the convolved particle distribution can only be recovered within this region. A carefully designed padding strategy with appropriate boundary conditions could potentially mitigate this issue.

An initially confusing observation is the discrepancy in the optimal model parameters for the nanoparticles, which were different in the simulation of convolution kernels than in the simulation of the system matrices.
However, since the inevitable removal of some frequency components in DCR leads to the loss of some low-frequency components, a larger nanoparticle core diameter was found to be optimal for DCR compared to system matrix reconstruction. As the core diameter increases, the essential support of the corresponding convolution kernel decreases, meaning that larger core diameters correspond to reduced blurring caused by the elimination of low spatial frequencies.

The results further demonstrate that the rank-$p$ approximation of the convolution kernels is efficient and can achieve arbitrary precision by adjusting the rank. Since the Langevin convolution kernel is nearly obtained for rank $p=1$, this approximation can also be interpreted as a generalization: For rank $p=1$, the solution closely follows the equilibrium model, while increasing $p$ progressively refines the result toward the equilibrium model with anisotropy.
By evaluating the fast operator within an iterative optimization algorithm, the proposed approach achieves a significantly faster runtime compared to the conventional method, which requires setting up a convolution matrix to account for local variations in the convolution kernels. This advantage becomes more pronounced as the FOV resolution increases, since the runtime scales as $\mathcal{O}(N \log (N))$ rather than $\mathcal{O}(N^2)$. In principle, the DCR method can also be applied to the reconstruction of three-dimensional data, where its superior runtime scaling with respect to the number of voxels becomes particularly beneficial.
In contrast to the proposed iterative approach, representing the deconvolution problem using a convolution matrix allows for a direct solution by inverting the matrix. However, as demonstrated in~\cite{droigk2023adaption}, this approach is particularly disadvantageous in terms of runtime. Moreover, employing iterative optimization algorithms enables the use of regularization terms beyond Tikhonov regularization, making an iterative solution approach preferable in general.
Furthermore, the rank-$p$ approximation of the convolution operator eliminates the need to construct a convolution matrix, offering significant advantages in terms of memory consumption. Similarly to the runtime benefits, this advantage becomes especially relevant for three-dimensional or high-resolution two-dimensional reconstructions. 

Owing to its improved reconstruction quality -- now also validated on real measurement data -- and its simultaneous computational acceleration, the DCR-EQANIS presented here represents an attractive alternative to state-of-the-art reconstruction using a measured system matrix. While achieving comparable image quality, the DCR eliminates the need for time-consuming system matrix calibration measurements. Its resolution flexibility, combined with low computational and memory requirements, further enhances its practicality.

In comparison to model-based system matrix reconstructions, DCR also benefits from its system-matrix-free nature, offering both superior image quality and operational advantages. Moreover, for other model-based reconstruction approaches, successful application to real measurement data using FFP-Lissajous trajectories has not yet been demonstrated, leaving their practical applicability in this context uncertain and precluding direct comparison.

\section{Conclusion}

This work assesses the advantages of the adapted DCR-EQANIS model compared to the classical DCR-EQ model using experimental measurement data. Six different phantoms were employed to evaluate the reconstruction quality of both approaches. The results demonstrate that the incorporation of magnetic anisotropies by the DCR-EQANIS model significantly improves the reconstruction accuracy, even when applied to real experimental data.

To mitigate the increased computational complexity of DCR-EQANIS, an efficient approximation of the spatially varying convolution was introduced. This approach allows for a flexible trade-off between accuracy and computational effort by performing the convolution operations and their adjoint operators in $\mathcal{O}(N \log N)$. Additionally, the compact storage of convolution kernels significantly reduces memory requirements, as the explicit construction of a convolution matrix is no longer necessary.

The use of an iterative optimization algorithm further enhances the speed of the operator evaluations during deconvolution, while also enabling the incorporation of alternative regularization techniques beyond the standard Tikhonov regularization.

While this work was restricted on two-dimensional experiments, the improvements in speed and memory consumption offer significant potential for application to large three-dimensional datasets. Future research should explore the extension of this approach to three-dimensional reconstructions, where the advantages of reduced computational cost and memory usage will be even more pronounced.

Overall, the DCR-EQANIS approach presented here demonstrates improved image quality and significantly accelerated computation times, establishing it as a practical and efficient model-based reconstruction method. Among model-based approaches developed for FFP-Lissajous trajectories, it currently offers the most favorable combination of performance and flexibility. Given its system-matrix-free nature and reconstruction quality comparable to state-of-the-art methods based on measured system matrices, DCR-EQANIS may even hold the potential to replace such methods in specific application scenarios.

\balance 
\bibliographystyle{IEEEtran}
\bibliography{ref}

\end{document}